\documentclass[aps,pra,twocolumn,showpacs]{revtex4}
\usepackage{amssymb}
\usepackage{amsmath}
\usepackage{graphicx}
\usepackage{epsfig}
\usepackage{subfigure}
\usepackage{color}
\begin{document}

\title{ Quantum correlations dynamics under different non-markovian environmental models}
\author{Ying-Jie Zhang,$^{1}$ }
\thanks{Email: yingjiezhang2007@163.com}
\author{Wei Han,$^{1}$ Chuan-Jia Shan,$^{2}$ Yun-Jie Xia$^{1}$}
\thanks{Email: yjxia@mail.qfnu.edu.cn}
\affiliation {$^{1}$Shandong Provincial Key Laboratory of Laser
Polarization and Information Technology, Department of Physics, Qufu
Normal University, Qufu 273165, China\\
$^{2}$College of Physics and Electronic Science, Hubei Normal
University, Huangshi 435002, China }
\date{\today}

\begin{abstract}
  We investigate the roles of different
environmental models on quantum correlation dynamics of two-qubit
composite system interacting with two independent environments. The
most common environmental models (the single-Lorentzian model, the
squared-Lorentzian model, the two-Lorentzian model and band-gap
model) are analyzed. First, we note that for the weak coupling
regime, the monotonous decay speed of the quantum correlation is
mainly determined by the spectral density functions of these
different environments. Then, by considering the strong coupling
regime we find that, contrary to what is stated in the weak coupling
regime, the dynamics of quantum correlation depends on the
non-Markovianity of the environmental models, and is independent of
the environmental spectrum density functions.
\end{abstract}

\pacs {03.67.Mn, 03.65.Yz, 42.50.-p, 71.55.Jv}

\maketitle
\section{\textbf{INTRODUCTION}}
\indent Until recently a lot of interest has been devoted to the
definition and understanding the quantum aspects of correlation in a
composite system. The discovery that mixed separable (unentangled)
states can have nonclassical correlation [1-4] and such states
provide computational speedup compared to classical states in some
quantum computation models [4,5]. In order to quantify the
quantumness of the correlation contained in a bipartite quantum
state Olliver and Zurek [3] proposed a measure for quantum
correlation known as quantum discord (QD) and based on a distinction
between quantum information theory and classical information theory.
A recent result that almost all quantum states have a nonvanishing
QD [6] shows up the relavance of studying such correlation.\\
\indent Besides the quantification of quantum correlations, another
important problem is the behavior of these correlations under the
action of decoherence. The phenomenon, caused by the injection of
noise into the system and arising from its inevitable interaction
with the surrounding environment, is responsible for the loss of
quantum coherence initially present in the system. Recently it was
noted that the dynamical behaviors of QD in the presence of the
Markovian [7,8] decoherence decay exponentially in
 time and vanish only asymptotically [9,10], contrary to the
 entanglement dynamics where sudden death may occur [11-18]. In these
 above studies, the quantumness of correlation is more robust to the
 action of the environment than the entanglement itself. In
 particular, Refs. [19,20] have discovered that the QD can
 be completely unaffected by Markovian depolarizing channels or non-Markovian depolarizing channels for long intervals of
 time, and this phenomenon has been observed experimentally [21]. As Refs. [19-21], it is of interest to find a certain environmental model
that the quantum correlation
can be unaffected by decoherence as much as possible.\\
\indent In this article, we will concentrate on the question: what
kind of local environmental model can make the initial quantum
correlation more robust in the dynamics process? We consider a
noninteracting two-qubit system under the influence of two
independent environments. The most common environmental models (the
single-Lorentzian model, the squared-Lorentzian model, the
two-Lorentzian model and band-gap model) are studied. By analytical
and numerical analysis we find that, for the weak coupling regime,
the monotonous decay speed of the two-qubit QD is mainly determined
by the spectral density functions of these different environments.
The two-qubit QD in the single-Lorentzian (band-gap) environment is
more robust than in the squared-Lorentzian (two-Lorentzian)
environment under the resonant and near resonant conditions. While
for the far off-resonant condition the two-qubit QD in the
single-Lorentzian (band-gap) environment decreases much more faster
than in the squared-Lorentzian (two-Lorentzian) environment.
However, by considering the strong coupling regime we find that, the
two-qubit QD is more robust in the squared-Lorentzian
(two-Lorentzian) environment than in the single-Lorentzian
(band-gap) environment, either under the near resonant or far
off-resonant condition. In this case, the dynamics of QD mainly
depends on the non-Markovianity of the environmental models, and is
independent of the environmental spectrum density functions.
\section{\textbf{Theoretical model and dynamics of two-qubit system}}
\indent Considering a model consisting of two qubits $A$ and $B$,
each interacting with a zero-temperature bosonic environment,
denoted $a$ and $b$, respectively, we assume that each
qubit-environment system is isolated and the environments are
initially in the vacuum state while two qubits are initially in an
quantum correlated state. A specific system which consists of two
independent two-level atoms interacting with an multi-mode
environment respectively has been chosen in this paper. Since each
atom evolves independently, we can learn how to characterize the
evolution of the overall system from the atom-environment dynamics.
The interaction between an atom and an N-mode environment in the
rotating-wave approximation is given by $H^{j}_{0}+H^{j}_{int}$,
which, in the basis
$\{|gg\rangle,|ge\rangle,|eg\rangle,|ee\rangle\}$, reads
\begin{eqnarray}
\hat{H}^{j}_{0}&=&\omega_{0}\hat{\sigma}^{j}_{+}\hat{\sigma}^{j}_{-}+\sum_{k=1}^{N}\omega_{k}a^{\dag}_{k}a_{k},\label{00}\\
\hat{H}^{j}_{int}&=&\sum_{k=1}^{N}g_{k}(\hat{\sigma}^{j}_{+}a_{k}+\hat{\sigma}^{j}_{-}a^{\dag}_{k}),\label{01}
\end{eqnarray}
here $a^{\dag}_{k}$, $a_{k}$ are the creation and annihilation
operators of quanta of the environment ($a$ or $b$),
$\hat{\sigma}^{j}_{+}=|e_{j}{\rangle}{\langle}g_{j}|$,
$\hat{\sigma}^{j}_{-}=|g_{j}{\rangle}{\langle}e_{j}|$ and
$\omega_{j}$ are the inversion operators and transition frequency of
the $j$-th atom (j=$A$, $B$ and here
$\omega_{A}=\omega_{B}=\omega_{0}$ ); $\omega_{k}$ and $g_{k}$ are
the frequency of the mode $k$ of the environment and its coupling
strength with the atom. To illustrate the roles of the different
environmental models on quantum correlation dynamics of two atoms,
we assume that two atoms interact off-resonantly with their
structured environment, whose spectral density function $D(\omega)$
provides a complete characterization of the evolution for
single-Lorentzian, two-Lorentzian, band-gap and squared-Lorentzian
environments.\\
\indent In order to find the atom-environment dynamics, we solve the
master equation by using the pseudomode approach [22,23]. This exact
master equation describes the coherent interaction between the atom
and the pseudomodes in the presence of the decay of the pseudomodes
due to the interaction with a Markovian reservoir [24]. The number
of the pseudomodes relies on the shape of the environmemt spectral
density function. $(a)$ For the single-Lorentzian environmental
model
$D(\omega)=\frac{\Gamma}{(\omega-\omega_{c})^{2}+(\Gamma/2)^{2}}$,
there has only one pole in the lower half complex plane, the atom
interacts with one pseudomode which leaks into a Markovian
environment. So the exact dynamics of the atom interacting with a
single-Lorentzian structured environment is contained in the
following pseudomode master equation
\begin{eqnarray}
\frac{d\rho}{dt}=-i[H^{j},\rho]-\frac{\Gamma}{2}[a^{\dag}a\rho-2a{\rho}a^{\dag}+{\rho}a^{\dag}a],\label{02}
\end{eqnarray}
where
\begin{eqnarray}
H^{j}=\omega_{0}\sigma^{j}_{+}\sigma^{j}_{-}+\omega_{c}a^{\dag}a+\Omega(\sigma^{j}_{+}a+\sigma^{j}_{-}a^{\dag}),\label{03}
\end{eqnarray}
with $\rho$ is the density operator for the $j$-th atom and the
pseudomode of the structured reservoir, $a$ and $a^{\dag}$ are the
annihilation and creation operators of the pseudomode. The constants
$\omega_{c}$ and $\Gamma$ are, respectively, the oscillation
frequency and the decay rate of the pseudomode and they depend on
the position of the pole $z\equiv\omega_{c}-i\Gamma/2$. The $j$-th
atom interacts coherently with the pseudomode (the strength
of the coupling $\Omega$).\\
\indent $(b)$ According to the two-Lorentzian environmental model,
the environment spectral density function is simply a sum of two
Lorentzian functions
$D(\omega)=W_{1}\frac{\Gamma_{1}}{(\omega-\omega_{c})^{2}+(\Gamma_{1}/2)^{2}}+W_{2}\frac{\Gamma_{2}}{(\omega-\omega_{c})^{2}+(\Gamma_{2}/2)^{2}}$,
where the weights of the two Lorentzians are such that
$W_{1}+W_{2}=1$. There are two poles in the lower half complex
plane, the atom interacts with two pseudomodes ($a_{1}$ and $a_{2}$)
which leak into a Markovian environment ($\Gamma_{1}$ and
$\Gamma_{2}$ are the decay rates), respectively. This time the poles
are located at $z_{1}=\omega_{c}-i\Gamma_{1}/2$ and
$z_{2}=\omega_{c}-i\Gamma_{2}/2$, so the exact master equation for
the atom-environment dynamics in the two-Lorentzian environmental
model can be written
\begin{eqnarray}
\frac{d\rho}{dt}=&-&i[H^{j},\rho]-\frac{\Gamma_{1}}{2}(a^{\dag}_{1}a_{1}\rho-2a_{1}{\rho}a^{\dag}_{1}+{\rho}a^{\dag}_{1}a_{1})\nonumber\\
&-&\frac{\Gamma_{2}}{2}(a^{\dag}_{2}a_{2}\rho-2a_{2}{\rho}a^{\dag}_{2}+{\rho}a^{\dag}_{2}a_{2}),\label{04}
\end{eqnarray}
and here
\begin{eqnarray}
H^{j}=&\omega_{0}&\sigma^{j}_{+}\sigma^{j}_{-}+\omega_{c}a^{\dag}_{1}a_{1}+\omega_{c}a^{\dag}_{2}a_{2}+\Omega\sqrt{W_{1}}(\sigma^{j}_{+}a_{1}\nonumber\\
&+&\sigma^{j}_{-}a^{\dag}_{1})+\Omega\sqrt{W_{2}}(\sigma^{j}_{+}a_{2}+\sigma^{j}_{-}a^{\dag}_{2}).\label{05}
\end{eqnarray}
\indent $(c)$ Next we give an idealized model of a band gap (or
photon density of states gap)
$D(\omega)=\frac{W_{1}\Gamma_{1}}{(\omega-\omega_{c})^{2}+(\frac{\Gamma_{1}}{2})^{2}}-\frac{W_{2}\Gamma_{2}}{(\omega-\omega_{c})^{2}+(\frac{\Gamma_{2}}{2})^{2}}$
in which both Lorentzians are centered at the same frequency, the
second is given a negative weight, and the weights of the two
Lorentzians are such that $W_{1}-W_{2}=1$ and
$\Gamma_{2}<\Gamma_{1}$ ensure positivity of $D(\omega)$. There also
have two poles in the lower half complex plane as the two-Lorentzian
model, the two poles are located at $\omega_{c}-i\Gamma_{1}/2$ and
$\omega_{c}-i\Gamma_{2}/2$, so there are also two pseudomodes
$a_{1}$ and $a_{2}$ with deacy rates
$\Gamma'_{1}=W_{1}\Gamma_{2}-W_{2}\Gamma_{1}$ and
$\Gamma'_{2}=W_{1}\Gamma_{1}-W_{2}\Gamma_{2}$ respectively. The
$j$-th atom does not couple to the first pseudomode $a_{1}$ at all,
it only interacts coherently with the second pseudomode $a_{2}$ (the
strength of the coupling $\Omega$) which is in turn coupled to the
first one (the strength of the coupling
$V=\sqrt{W_{1}W_{2}}(\Gamma_{1}-\Gamma_{2})/2$), and both
pseudomodes are leaking into independent Markovian environments. The
exact pseudomode master equation associated with the band-gap model
is given by
\begin{eqnarray}
\frac{d\rho}{dt}=&-&i[H^{j},\rho]-\frac{\Gamma'_{1}}{2}[a^{\dag}_{1}a_{1}\rho-2a_{1}{\rho}a^{\dag}_{1}+{\rho}a^{\dag}_{1}a_{1}]\nonumber\\
&-&\frac{\Gamma'_{2}}{2}[a^{\dag}_{2}a_{2}\rho-2a_{2}{\rho}a^{\dag}_{2}+{\rho}a^{\dag}_{2}a_{2}],\label{06}
\end{eqnarray}
where
\begin{eqnarray}
H^{j}&=&\omega_{0}\sigma^{j}_{+}\sigma^{j}_{-}+\omega_{c}a^{\dag}_{1}a_{1}+\omega_{c}a^{\dag}_{2}a_{2}+\Omega(a^{\dag}_{2}\sigma^{j}_{-}\nonumber\\
&+&a_{2}\sigma^{j}_{+})+V(a^{\dag}_{1}a_{2}+a_{1}a^{\dag}_{2}).
\label{07}
\end{eqnarray}
\indent $(d)$ The environment spectral density function of the
squared-Lorentzian model is
$D(\omega)=\frac{\Gamma^{3}/2}{[(\omega-\omega_{c})^{2}+(\Gamma/2)^{2}]^{2}}$,
for which we will find that there exist two pseudomodes  $a_{1}$ and
$a_{2}$, and the $j$-th atom only couples to the second pseudomode
$a_{2}$ (the coupling constant $\Omega$) which interacts with the
first pseudomode $a_{1}$ (the strength of the coupling
$V=\Gamma/2$). Different from the band-gap model, only the first
pseudomode will show any decay to the Markovian environment with
decay rate $\Gamma$, the second pseudomode which is directly coupled
to the $j$-th atom does not decay in this model. So the dynamics of
the $j$-th atom and two pseudomodes obey the following master
equation
\begin{eqnarray}
\frac{d\rho}{dt}=-i[H^{j},\rho]-\Gamma[a^{\dag}_{1}a_{1}\rho-2a_{1}{\rho}a^{\dag}_{1}+{\rho}a^{\dag}_{1}a_{1}],\label{08}
\end{eqnarray}
with the Hamiltonian
\begin{eqnarray}
H^{j}&=&\omega_{0}\sigma^{j}_{+}\sigma^{j}_{-}+\omega_{c}a^{\dag}_{1}a_{1}+\omega_{c}a^{\dag}_{2}a_{2}+\Omega(a^{\dag}_{2}\sigma^{j}_{-}\nonumber\\
&+&a_{2}\sigma^{j}_{+})+V(a^{\dag}_{1}a_{2}+a_{1}a^{\dag}_{2}).\label{09}
\end{eqnarray}
\indent In order to analyze the roles of the different environmental
models on quantum correlation dynamics of two atoms, we consider the
above four environmental models, respectively, $i.e.,$ the
single-Lorentzian model, two-Lorentzian model, band-gap model and
squared-Lorentzian model. According to the above analysis, the
spectral density functions of single-Lorentzian model and
squared-Lorentzian model have a same parameter $\Gamma$, and the
two-Lorentzian model and band-gap model both contain two
Lorentzians, the same parameters ($W_{1}$, $W_{2}$, $\Gamma_{1}$ and
$\Gamma_{2}$) appear in the spectral density functions of them. So
in this paper we will mainly compare the difference in quantum
correlation dynamics of two atoms between the single-Lorentzian
model and squared-Lorentzian model, as well as bewteen the
two-Lorentzian model and band-gap model. For an initial state of the
total system $\rho(0)_{ABab}=|\Psi\rangle\langle\Psi|$, with
$|\Psi\rangle=(\cos\theta|gg\rangle_{AB}+\sin\theta|ee\rangle_{AB})\otimes|\bar{0}\rangle_{a}|\bar{0}\rangle_{b}$,
and here $\theta\in[0,\pi]$, $|e\rangle$ and $|g\rangle$ are the
excited state and ground state of atoms,
$|\bar{0}\rangle_{a,b}=\prod_{k=1}^{N}|0_{k}\rangle_{a,b}$ is the
vacuum state of the environment $a,b$. Then the evolutional density
matrix $\rho(t)$ of the total system in different environmental
models can be acquired respectively by solving the above master
equations (from Eqs. (\ref{02}) to (\ref{09})). Tracing out the
pseudomode degree of freedom, we obtain the reduced density matrix
$\rho_{AB}(t)$ of the
atomic system in these four different environmental models.\\
\indent  The measure of total quantum correlations used here is the
quantum discord (QD) [3]. In all cases investigated in this paper
the reduced density matrix for the atomic system $\rho_{AB}(t)$ in
the basis $\{|gg\rangle,|ge\rangle,|eg\rangle,|ee\rangle\}$ has an
$X$ structure defined by its elements
$\rho_{12}=\rho_{13}=\rho_{24}=\rho_{34}=0$, $\rho_{22}=\rho_{33}$
and $\rho_{14}=\rho^{*}_{41}$. For this $X$ class of density matrix,
QD can be calculated analytically [25]:
$QD(\rho_{AB})=S(\rho_{B})+\sum^{3}_{j=0}\lambda_{j}\log_{2}\lambda_{j}+\min_{\{B_{i}\}}[S(\rho|\{B_{i}\})]$,
where $\lambda_{j}$ is the $j$-th eigenvalue of the density matrix
$\rho_{AB}(t)$. Here $S(\rho_{B})$ denotes the von Neumann entropy
of $\rho_{B}=Tr_{A}\rho_{AB}$ and $S(\rho|\{B_{i}\})$ is the quantum
conditional entropy with respect to a von Neumann measurement
$\{B_{i}\}$ for subsystem $B$.\\
\begin{figure}[tbp]
\includegraphics[scale=0.8]{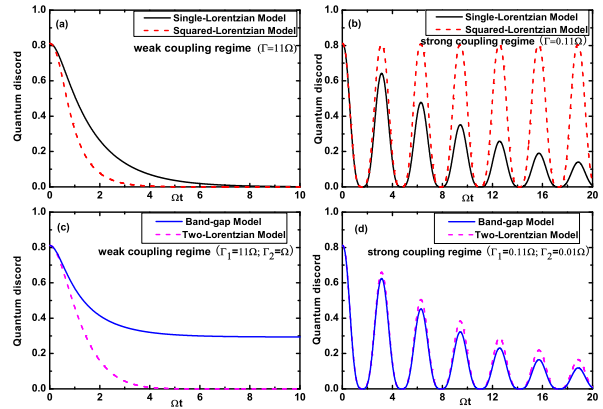}
\caption{\label{fig1}(Color online) Time evolution of the atomic QD
as a function of the dimensionless quantity ${\Omega}t$ under the
atom-pseudomode resonant condition
($\Delta=\omega_{c}-\omega_{0}=0$), with $\theta=\pi/3$. (a) and (b)
the single-Lorentzian and squared-Lorentzian as the environmental
models; (c) and (d) the two-Lorentzian and band-gap as the
environmental models.}
\end{figure}

\section{\textbf{Numerical results and discussions}}
\indent In Figs. $1(a)$ and $1(b)$, by considering the
atom-pseudomode resonant condition
($\Delta=\omega_{c}-\omega_{0}=0$) and choosing the
single-Lorentzian
 model and squared-Lorentzian model as the environmental spectral density
 functions, we plot the time evolution of QD for two
 qubits as function of the dimensionless quantity ${\Omega}t$ in the
 weak coupling regime and the strong coupling
 regime, with $\theta=\pi/3$. In the weak coupling regime QD decays only
asymptotically to zero while in the strong coupling regime the
atomic QD presents damped oscillations. A comparison between the
dark solid curve and the red dashed curve in Figs. $1(a)$ and $1(b)$
reveals that for the weak coupling regime, the atomic QD due to the
single-Lorentzian environmental model is more robust than the case
of the squared-Lorentzian model, but in the strong coupling regime,
the atomic QD attenuates more slowly in the case of the
squared-Lorentzian model than the single-Lorentzian model. Through
comparing the atomic QD dynamics in the two-Lorentzian environmental
model and band-gap environmental model (as shown in Figs. $1(c)$ and
$1(d)$), we acquire that in the weak coupling regime, the QD can
decrease to a long-time asymptotic value in the band-gap model while
for the case of the two-Lorentzian model the QD can reduce
eventually to zero. However, according to the strong coupling
regime, the atomic QD is more robust in the case of the
two-Lorentzian model than the band-gap model.\\
\indent Then, in order to investigating the effects of different
environmental models on the atomic QD under the atom-pseudomode near
resonant and far off-resonant conditions, we analyze the evolution
behavior of the QD in the weak coupling regime, by the comparison of
two cases: near resonance condition ($\Delta=0.2\Omega$) and far
off-resonance condition ($\Delta=8\Omega$), with $\theta=\pi/3$. For
the case of near resonance, as shown in Figs. $2(a)$ and $2(c)$, one
could find that the atomic QD in the single-Lorentzian (band-gap)
environment is more robust than in the squared-Lorentzian
(two-Lorentzian) environment. However, an opposite result that the
atomic QD in the single-Lorentzian (band-gap) environment decreases
much more faster than in the squared-Lorentzian (two-Lorentzian)
environment are obtained for the far off-resonant condition, clearly
seen in Figs. $2(b)$ and $2(d)$. What is the physics behind the
phenomena? In this part we try to give an enlightening discussion
for this problem based on these environmental models. Let us review
the spectrum density functions of these models, as shown in Fig.
$3$. The center part of the spectrum density function of the
single-Lorentzian (band-gap) environment is much smaller than of the
squared-Lorentzian (two-Lorentzian) environment. In contrast, the
parts which are far from the center are larger in the
single-Lorentzian (band-gap) environment than in the
squared-Lorentzian (two-Lorentzian) environment. Thus, to determine
in which environmental model the atomic QD is more robust in the
weak coupling regime, we can compare the spectral density functions
of these different environments: the decay behavior of the atomic QD
is determined by the modes of the spectrum which are resonant with
the atoms: the monotonous decay
speed of the QD decreases as the density of these modes decreases.\\
\begin{figure}
\includegraphics[scale=0.8]{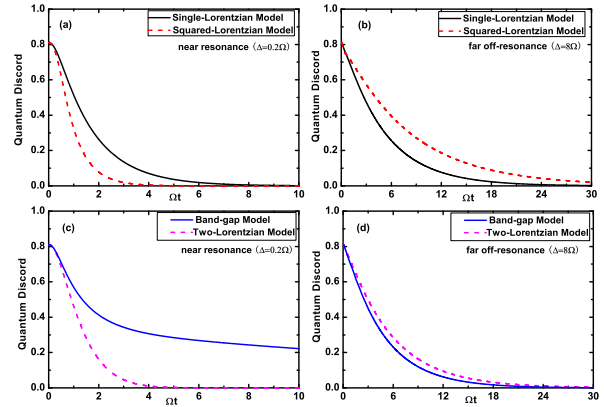}
\caption{\label{fig2}(Color online) Time evolution of the atomic QD
as a function of the dimensionless quantity ${\Omega}t$ under the
atom-pseudomode near resonance regime and far off-resonance regime,
with $\theta=\pi/3$. (a) and (b) the single-Lorentzian and
squared-Lorentzian as the environmental models, with
$\Gamma=11\Omega$; (c) and (d) the two-Lorentzian and band-gap as
the environmental models, with $\Gamma_{1}=11\Omega$,
$\Gamma_{2}=\Omega$.}
\end{figure}
\begin{figure}
\includegraphics[scale=0.9]{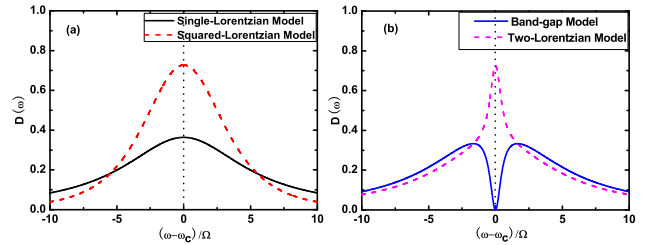}
\caption{\label{fig3}The density of the spectrum $D(\omega)$ as a
function of the dimensionless quantity $(\omega-\omega_{c})$ in the
weak-coupling regime. (a) for the single-Lorentzian
 and squared-Lorentzian as the environmental models, with $\Gamma=11\Omega$; (b) for the two-Lorentzian
 and band-gap as the environmental models, with $\Gamma_{1}=11\Omega$, $\Gamma_{2}=\Omega$.}
\end{figure}
\indent In this paper, we also understand the influences of
different environments on the atomic QD in the strong coupling
regimes which satisfy $\Gamma=0.11\Omega$ in the single-Lorentzian
environment and squared-Lorentzian environment, and
$\Gamma_{1}=0.11\Omega$, $\Gamma_{2}=0.01\Omega$ in the
two-Lorentzian environment and band-gap environment. In  Fig. $4$,
we acquire that the periodically oscillating decay speed of the
atomic QD in the squared-Lorentzian (two-Lorentzian) environment is
slower than in the single-Lorentzian (band-gap) environment, either
under the atom-pseudomode near resonant or far off-resonant
condition. That is to say, in the strong coupling regime the QD is
more robust in the squared-Lorentzian (two-Lorentzian) environment
than in the single-Lorentzian (band-gap) environment. In what
follows, we will give a simple interpretation for why this finding
in the strong regime is different from the results in the weak
coupling regime. First, taking the spectrum density function
$D(\omega)$ of the above four environmental models into account in
the strong regime, we note that the discrepancy among them is very
minor, as shown in Fig. $5$. So from the spectrum density function
to give a construction is not feasible.\\
\begin{figure}
\includegraphics[scale=0.8]{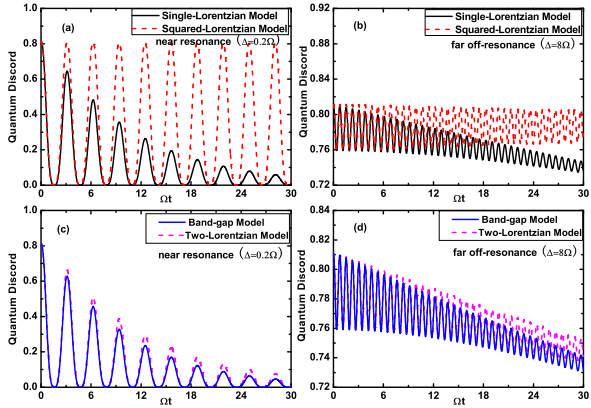}
\caption{\label{fig4}(Color online) Time evolution of the atomic QD
as a function of the dimensionless quantity ${\Omega}t$ under the
atom-pseudomode near resonance regime and far off-resonance regime,
with $\theta=\pi/3$. (a) and (b) the single-Lorentzian and
squared-Lorentzian as the environmental models, with
$\Gamma=0.11\Omega$;
 (c) and (d) the two-Lorentzian and band-gap as the environmental models, with $\Gamma_{1}=0.11\Omega$, $\Gamma_{2}=0.01\Omega$.}
\end{figure}
\begin{figure}
\includegraphics[scale=0.6]{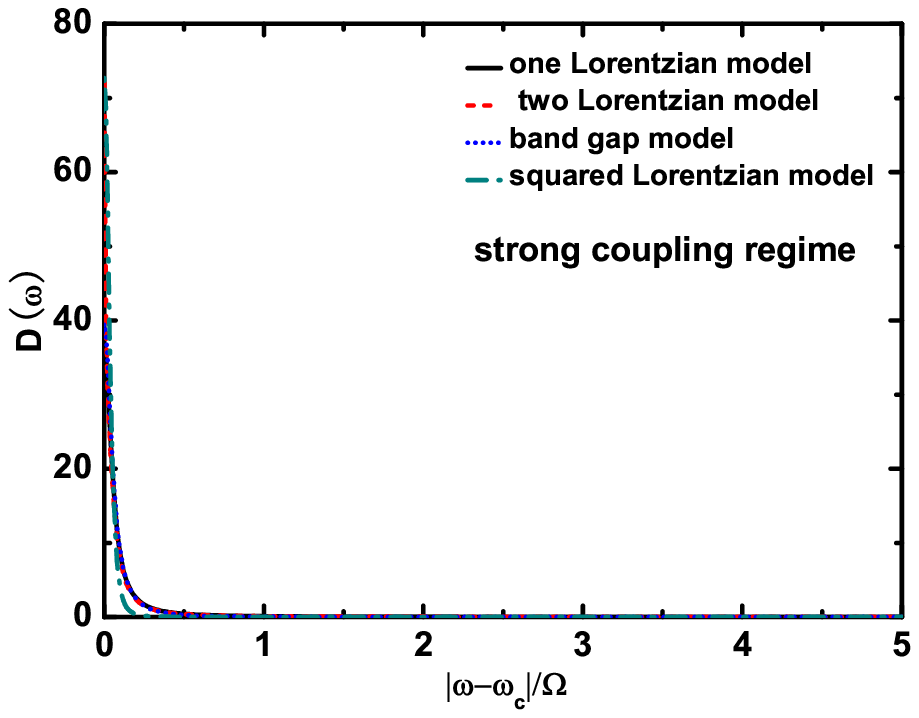}
\caption{\label{fig5}The density of the spectrum $D(\omega)$ as a
function of the dimensionless quantity $(\omega-\omega_{c})$ in the
strong-coupling regime, for the single-Lorentzian
 and squared-Lorentzian as the environmental models, with $\Gamma=0.11\Omega$; and for the two-Lorentzian
 and band-gap as the environmental models, with $\Gamma_{1}=0.11\Omega$, $\Gamma_{2}=0.01\Omega$.}
\end{figure}
\begin{figure}
\includegraphics[scale=1.0]{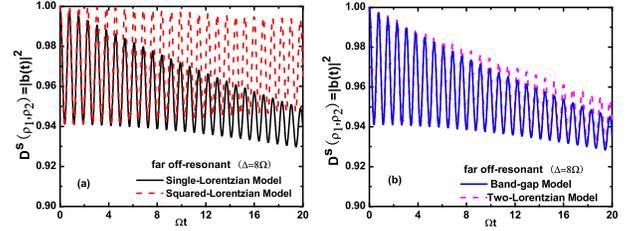}
\caption{\label{fig6}(Color online) Time evolution of the trace
distance $D^{S}(\rho_{1},\rho_{2})$ as a function of the
dimensionless quantity ${\Omega}t$ under the far off-resonance
regime, with $\rho_{1}(0)=|e\rangle{\langle}e|$ and
$\rho_{2}(0)=|g\rangle{\langle}g|$. (a) for the single-Lorentzian
and squared-Lorentzian as the environmental models, with
$\Gamma=0.11\Omega$;
 (b) for the two-Lorentzian and band-gap as the environmental models, with
$\Gamma_{1}=0.11\Omega$, $\Gamma_{2}=0.01\Omega$.}
\end{figure}
\indent However, according to the previous works [26-28], we know
that there exists the non-Markovianity of environment in the strong
coupling regime, and the non-Markovian effect of environment can
play an important role on the dynamics of the qubits system.
Therefore, we will show the degree of the non-Markovian behavior of
the dynamics processes in these different environmental models. In
Ref. [26], Breuer $et$ $al.$ suggest definition a measure $N(\Phi)$
for the non-Markovianity of the quantum process $\Phi(t)$ by means
of the relation
$N(\Phi)=\max_{\rho_{1,2}(0)}\sum_{i}[D(\rho_{1}(b_{i}),\rho_{2}(b_{i}))-D(\rho_{1}(a_{i}),\rho_{2}(a_{i}))]$.
To calculate this quantity one first determines the total growth of
the trace distance over each time interval $(a_{i},b_{i})$ and sums
up the contributions of all intervals. Then $N(\Phi)$ can be
obtained by determining the maximum over all pairs of initial
states. Taking the far off-resonance as an example, the analytical
expression of the trace distance in the atom-environment dynamics
process is $D^{S}(\rho_{1},\rho_{2})=|b(t)|^{2}$, here with $b(t)$
represents the amplitude damping of the excited state $|e\rangle$,
and the pair of initial states $\rho_{1}(0)=|e\rangle{\langle}e|$
and $\rho_{2}(0)=|g\rangle{\langle}g|$ which optimize the total
increase of $D^{S}(\rho_{1},\rho_{2})$. Thus, we can qualitatively
and intuitively compare the non-Markovianity due to the different
environment models by the time evolution of the trace distance. Fig.
$6$ shows the trace distance $D^{S}(\rho_{1},\rho_{2})$ as a
function of ${\Omega}t$ for $\Delta=8\Omega$, namely, the far
off-resonance regime. It is very interesting to note that the
amplitudes of $D^{S}(\rho_{1},\rho_{2})$ caused by
squared-Lorentzian (two-Lorentzian) environment are much wider than
those caused by the single-Lorentzian (band-gap) environment. In
other words, the non-Markovianity of the squared-Lorentzian
(two-Lorentzian) environment is much stronger than the
single-Lorentzian (band-gap) environment. This finding leads to a
clear interpretation for the result obtained by Fig. $4$: the atomic
QD in the strong coupling regime is determined by the different
degree of environmental non-Markovianity, and is independent of the
spectrum density function $D(\omega)$.\\
\indent In conclusion, we have studied the quantum correlation
dynamics in the different decoherence environments, and considered a
two-atom system interacting with two local, independent
environments, modeling several common noise sources: the
single-Lorentzian model, the squared-Lorentzian model, the
two-Lorentzian model and band-gap model. For the weak coupling
regime, it is clear to realize that the atomic QD in the
single-Lorentzian (band-gap) environment is more robust than in the
squared-Lorentzian (two-Lorentzian) environment under the resonant
and near resonant conditions. But for the far off-resonant condition
the opposite result shows that the atomic QD in the
single-Lorentzian (band-gap) environment decreases much more faster
than in the squared-Lorentzian (two-Lorentzian) environment.
However, for the strong coupling regime, the atomic QD is more
robust in the squared-Lorentzian (two-Lorentzian) environment than
in the single-Lorentzian (band-gap) environment, either under the
atom-pseudomode near resonant or far off-resonant condition. Finally
we note that we study here only the two-atom system interacting with
their independent environments. An important future investigation
will be the study of the effects of these different environmental
models on the dynamics of the two-atom system under a common
environment, where quantum correlations can be created in the system
through nonlocal interactions mediated by the environment.
\\
\section{\textbf{Acknowledgments}}
\indent This work is supported by National Natural Science
Foundation of China under Grant Nos. 61178012 and 10947006, the
Specialized Research Fund for the Doctoral Program of
Higher Education under Grant No. 20093705110001 and the Research Funds from Qufu Normal University under Grant No. XJ201013.\\


\begin{thebibliography}{99}

\bibitem{[1]}V. Vedral, Phys. Rev. Lett. \textbf{90} 050401 (2003)\\
\bibitem{[2]}S. Luo, Phys. Rev. A \textbf{77} 042303 (2008)\\
\bibitem{[3]}H. Ollivier, and W. H. Zurek, Phys. Rev. Lett. \textbf{88} 017901 (2001)\\
\bibitem{[4]}A. Datta, A. Shaji, and C. Caves, Phys. Rev. Lett. \textbf{100} 050502 (2008)\\
\bibitem{[5]}B. P. Lanyon et al., Phys. Rev. Lett. \textbf{101} 200501 (2008)\\
\bibitem{[6]}A. Ferraro, L. Aolita, D. Cavalcanti, F. M. Cucchietti,
and A. Acin, e-print arXiv:quant- ph/0908.3157 (2009)\\
\bibitem{[7]}J. Maziero et al., Phys. Rev. A \textbf{80} 044102 (2009)\\
\bibitem{[8]}J. Maziero et al., Phys. Rev. A \textbf{81} 022116 (2010)\\
\bibitem{[9]}T. Werlang et al., Phys. Rev. A \textbf{80} 024103 (2009)\\
\bibitem{[10]}Y. J. Zhang, X. B. Zou, Y. J. Xia, and G. C. Guo, J. Phys. B \textbf{44} 035503 (2011)\\
\bibitem{[11]}Y. J. Zhang, X. B. Zou, Y. J. Xia, and G. C. Guo, Phys. Rev. A \textbf{82} 022108 (2010)\\
\bibitem{[12]}T. Yu, and J. H. Eberly, Phys. Rev. Lett. \textbf{93} 140404 (2004)\\
\bibitem{[13]}T. Yu, and J. H. Eberly, Science \textbf{323} 598 (2009).\\
\bibitem{[14]}B. Bellomo, R. L. Franco, and G. Compagno, Phys.
Rev. Lett. \textbf{99} 160502 (2007).\\
\bibitem{[15]}S. Maniscalco, F. Francia, R. L. Zaffino, N. L. Gullo,
and F. Plastina, Phys. Rev. Lett. \textbf{100} 090503 (2008).\\
\bibitem{[16]}B. Bellomo, R. L. Franco, S. Maniscalco, and G. Compagno, Phys. Rev. A \textbf{78} 060302(R) (2008).\\
\bibitem{[17]}Z. Ficek, and R. Tanas, Phys. Rev. A \textbf{77} 054301 (2008).\\
\bibitem{[18]}C. E. L\'{o}pez, G. Romero, F. Lastra, E. Solano, and J. C. Retamal, Phys. Rev. Lett. \textbf{101} 080503 (2008).\\
\bibitem{[19]}L. Mazzola, J. Piilo, and S. Maniscalco, Phys. Rev. Lett. \textbf{104} 200401 (2010).\\
\bibitem{[20]}L. Mazzola, J. Piilo, and S. Maniscalco, e-print
arXiv:quant-ph/1006.1805 (2010)\\
\bibitem{[21]}J. S. Xu, X. Y. Xu, C. F. Li, C. J. Zhang, X. B. Zou, and G. C. Guo, Nat. Commun. \textbf{1} 7 (2010)\\
\bibitem{[22]}B. M. Garraway, Phys. Rev. A \textbf{55} 4636 (1997)\\
\bibitem{[23]}B. M. Garraway, Phys. Rev. A \textbf{55} 2290 (1997)\\
\bibitem{[24]}L. Mazzola, S. Maniscalco, J. Piilo, K.-A. Suominen,
and B. M. Garraway, Phys. Rev. A \textbf{79} 042302 (2009);
\textbf{80},012104 (2009).\\
\bibitem{[25]}M. Ali, A. R. P. Rau, and G. Alber, Phys. Rev. A \textbf{81} 042105 (2010)\\
\bibitem{[26]}H. P. Breuer, E. M. Laine, and J. Piilo, Phys. Rev. Lett. \textbf{103} 210401 (2009)\\
\bibitem{[27]}Z. He, J. Zou, L. Li, and B. Shao, Phys. Rev. A \textbf{83} 012108 (2011)\\
\bibitem{[28]}A. Rivas, S. F. Huelga, and M. B. Plenio, Phys. Rev. Lett. \textbf{105} 050403 (2010)\\

\end{thebibliography}
\end{document}